\documentclass[a4paper,twocolumn]{article}
\usepackage{graphicx}
\usepackage{authblk}
\usepackage{cite}
\usepackage[english]{babel}
\usepackage{setspace}
\usepackage{relsize}
\usepackage{bm} 
\usepackage{amsmath}
\usepackage{amsfonts}
\title{Application of the theory of excitons to study of the positronium in matter. Optical transition during positronium formation in matter}
\author[]{M.~Pietrow\thanks{email: mrk@kft.umcs.lublin.pl}}
\affil[]{Institute of Physics, M. Curie-Sk{\l}odowska University, ul.~Pl.~M.~Curie-Sk{\l}odowskiej~1, 20-031 Lublin, Poland}
\begin{document}
\maketitle
\begin{figure}
\centering
\includegraphics[scale=0.35]{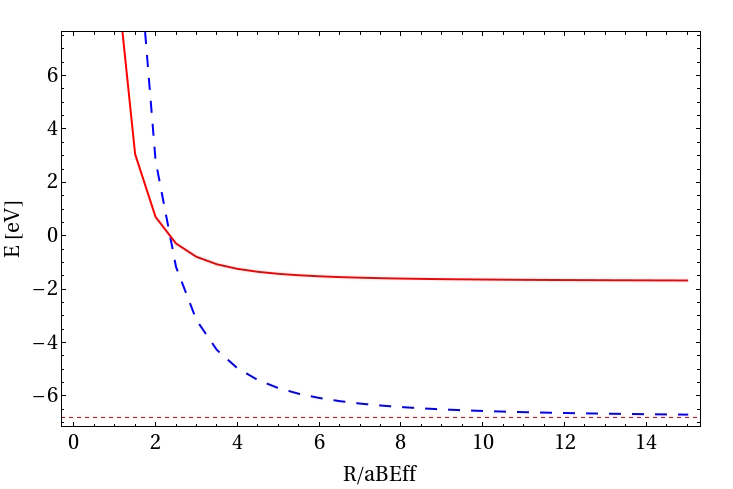}
\caption{An energy of $e^+$--$e^-$ pair in the free volume which is surrounded by a bulk with relative permittivity $\varepsilon_r$=1 (blue dashed line) and $\varepsilon_r$=2 (red plain line) as a function of radius of the free volume (in effective Bohr radii units). A finite potential step approach was used \cite{Kayanuma90}. The horizontal dashed line indicates the level of energy characteristic for Ps atom in the vacuum.}
\label{fig:Kayanuma90}
\end{figure}

\paragraph{Abstract}
Considerable similarity between positronium atom and an exciton in a quantum dot is indicated. Following this, it were applied the results from the theory of excitons to describe some aspects of formation of a positronium in matter. The possibility of photonic deexcitation during Ps formation was considered and the way of calculation of its probability was shown.\\
The photonic transitions speculated here, if detected, allows improving experimental studies of solid matter with positron techniques.
\paragraph{Introduction.}
An exciton is a pair of interacting particles formed by an electron ($e^-$) and a hole ($h$) in matter. The notion of the exciton allows interpretation of some spectroscopic shifts observed for both semiconductors and molecular crystals. The main concern of the contemporary science of excitons is to describe their properties in a confined space like quantum dots or quantum wires~\cite{Harrison05}. Excitons in quantum dots are regarded as a kind of artificial atoms produced mainly in semiconductor nano-domains. They are used as a source of light with controlled properties~\cite{Rasmussen13}.\\
A positronium (Ps) is a perishable atom made of an electron and a positron ($e^+$). Its properties in matter are used by positron techniques to study such properties of matter as its free volume size and bulk electron density~\cite{Schrader03}. The theory of positrons in matter is presented in detail in~\cite{Stepanov03}.\\
There is considerable similarity between the $e^+$--$e^-$ pair in the condensed matter and an exciton. This similarity seems not to have been presented yet in the literature in a wide extent. In the opinion of the Author, the quantum theory of excitons is a natural quantum language of description of a positron and positronium in matter. It can be applied as an extension of a recent formalism in the field of theoretical description of the positronium in matter. For example, the formalism borrowed from the exciton theory should facilitate systematic study of photonic and phononic transition in the case of positronium formation. Furthermore, it allows studying the $e^+$--$e^-$ pair in the trap of a free volume in a fully quantum way. Now, the Ps in a free volume is modeled as a quasi-classical undivided particle trapped in a single-particle quantum well~(for review, see~\cite{Goworek14}). There is no exhaustive theory yet how and where the energy excess released during the positronium formation is deposited (The Author's greatest concern are alkanes -- saturated hydrocarbons. Most approximations in this paper are made for them. For alkanes, this energy could be even ca.~3~eV \cite{Pietrow15}).
\\
The idea of linking positrons with excitons is not completely new. There are suggestions of such similarities when the Ps is formed at a surface of semiconductors~\cite{Cassidy12}. Furthermore, the positronium - exciton interaction problem was studied~\cite{Vorobiev76}. Nevertheless, none of these papers applies the similarity of the $e^+$--$e^-$ to the $h$--$e^-$ pair in the wide extent.\\
The novelty of this paper is to use an exciton concept to introduce a theory known from a solid state physics to support a theory of Ps formation near a free volumes. A part of this concept, i.e. interaction of Ps with phonons is already published~\cite{Bondarev98}. Here, the investigations of $e^+$--$e^-$ interaction with phonons is continued. Furthermore, it is argued here, in accordance with processes known for excitons, that the photonic emission is possible during the process of positronium formation in matter.
\paragraph{$\mathbf{e^+}$--$\mathbf{e^-}$ pair in a bulk material.}\label{Ps in bulk}
The similar Hamiltonian should be analyzed both for the exciton in a quantum well and for a $e^+$--$e^-$ pair in a material. Based on \cite{BasdevantDalibard02,Ivchenko05}, we can write the Schr\"{o}dinger equation as follows
\begin{equation}
\Big( \frac{\mathbf{p}_e^2}{2m_e}+\frac{\mathbf{p}_h^2}{2m_h}-\frac{ke^2}{r}+V_e+V_h\Big)\Psi(\vec{\mathbf{r}_e},\vec{\mathbf{r}_h})=E \Psi(\vec{\mathbf{r}_e},\vec{\mathbf{r}_h}),
\label{eq:Hamiltonian}
\end{equation}
where $k=1/(4\pi\epsilon_0\epsilon_r)$, $r$ -- a relative distance, $\mathbf{p}_i$ and $m_i$ -- momenta and masses, respectively. The index 'e' here is ascribed to an electron and 'h' to a positron or a hole. The potentials $V_i$ describe an interaction of an electron (hole) with the quantum wall. We assume that both the electron and the positron are placed in a mean attractive potential from the dipoles of the bulk \cite{Stepanov13,Pietrow15}. Suppose that $V_e=V_h=-V$. 
The electron's and positron's masses here, $m_e$ and $m_h$ respectively, denote the effective ones.\\
After transformation of coordinates and momenta to a center-of-mass coordinates,
the Hamiltonian splits into two parts:~a center-of-mass part and a part describing a relative motion.
The method for solution of this problem is analogous to that for a hydrogen atom. 
The energy spectrum has a band structure because of the continuous spectrum of $K$ and the discrete spectrum of the $n$ number,
\begin{equation}
E_{K,n}=-\Big(2V+\frac{\hbar^2 K^2}{2M}+ \frac{m_e k^2 e^4}{4 n^2 \hbar^2} \Big).
\label{eq:EnergyInBulk}
\end{equation}
According to the model described above, both the exciton in a bulk and the $e^+$-$e^-$ pair resemble a spherical atom moving through the bulk. The size of this atom depends strictly on the relative electric permittivity $\epsilon_r$ of the sample penetrated by positrons: the maximum of radial probability function $r^2 |R_{1,0}(r)|^2$ for $n$=1, $l$=0 for alkanes, where $\epsilon_r$=2 typically~\cite{crc_Handbook}, is ca. 2$\AA$ and for $\epsilon_r$=10 (alcohols) it is nearly 1~nm.\\
Some processes should be distinguished during a Ps formation. One of them is self-trapping of charges. The origin of this process is an interaction of $e^+$ or $e^-$ with surrounding molecules. According to my knowledge, both electron and a positron effective interaction with surrounding molecules is attractive one and the particles become bound (trapped). When they are far apart they loose their energy by an interaction with both optical and acoustic phonons. The time scale of starting of a process of polarons and phonons creation~\cite{Ueta86} is ca.~10$^{13}$~s (a characteristic time scale of lattice oscillations). When the pair approaches and can be considered as a charge-neutral object, a main consequence of the interaction is a local deformation of lattice made by localised charge state. In this way, dressed charges which form a quasi-free Ps (qf-Ps)~\cite{Stepanov03} do continue to move in a quasi-random regime and possibly form Ps.\\
One should take into account a process of self-trapping of qf-Ps. According to \cite{Bondarev98}, one should distinguish (in the case of alkali halides, at least) between a delocalised Ps state (which is identified with not-self-trapped state in a cited paper) and a localised one. The last one is a self-trapped state (with negative energy). A transition from a former to a latter one can be considered as a tunnelling process.
\paragraph{$\mathbf{e^+}$--$\mathbf{e^-}$ pair near a free volume.}
The $e^+$--$e^-$ pair near the free volume is of special concern in the Ps formation theory because it precedes the Ps formation. Oppositely to a general Ps formation theory~\cite{Stepanov03}, the exciton theory considers the $e$--$h$ pair in a dot's volume without an additional assumption that it has to form an atom-like structure. Keeping the analogy between the $e^+$-$e^-$ and the exciton near the free volume, according to~\cite{Kayanuma86,Kayanuma90,Pellegrini05,Tang05}, we can calculate the energy and the wavefunction of the pair in the following two opposite regimes. If the radius $r_0$ of the free volume is greater than the effective Bohr radius\footnote{$r_B^{\textrm{eff}}=\epsilon\hbar^2/(\mu e^2)$, where: $e$ - the electron charge, $\mu$ - a reduced mass and $\mu=(1/m_e+1/m_h)^{-1}$ whereas $\epsilon=4\pi\epsilon_0\epsilon_r$, where $\epsilon_0$, $\epsilon_r$ are the vacuum permittivity and the relative permittivity, respectively~\cite{Kayanuma86}.} $r_B^{\textrm{eff}}$ of the exciton (positronium atom), the pair should be regarded as a confined atom (weak-confinement regime \cite{Wang05}). On the contrary, if the radius of the exciton in the bulk exceeds the radius of the cavity, the interaction of individual particles with the wall dominates (the individual-particle confinement regime). In the latter case, the Coulomb interaction between these particles is regarded as perturbation. For $2\le r_{0} /r_B^{\textrm{eff}} \le 4$, which is possibly the most interesting here, the intermediate regime takes place.\\
Taking the analogy into account, it seems that considering the $e^+$--$e^-$ pair in relatively small free volumes as a positronium atom is a simplification. Figure \ref{fig:Kayanuma90}
shows (dashed line) the energy of the pair as a function of free volume radius expressed in units of the effective Bohr radii. The case was calculated by the Author for finite step potential~\cite{Kayanuma90}. Here the barrier height was established at 10.0~eV, $\varepsilon_r$ for the bulk surrounding the free volume was set at 1.0, whereas effective masses were $m_e$=$m_h$=$m_0$, where $m_0$ is an electron rest mass. The most important conclusion is that scarcely for the radius greater than 4$\cdot r_B^{\textrm{eff}}$ the energy of the pair does approach the energy of Ps in the vacuum. For radii with a lower value, the energy indicates a considerable influence of atom confinement. The model of the finite well shows that the spill of the wavefunction out of the free volume should not be neglected in this scale of the free volume size. Furthermore, if the bulk surrounding the free volume is able to polarise, the energy of the pair considerably changes due to the interaction with the bulk. A plain line in this figure shows the energy of the pair calculated for $\varepsilon_r$=2.
%
%
\paragraph{Phononic interaction with $\mathbf{e^+}$--$\mathbf{e^-}$ pair.}
One need to estimate the influence of interaction of $\mathbf{e^+}$--$\mathbf{e^-}$ with surrounding matter by phononic excitations. It seems that an qualitative approximation of it can be taken from a continuum model of self-trapping \cite{Williams93}. In this model, one take into account an interaction of an exciton with both optical and acoustic phonons. If the $\mathbf{e^+}$--$\mathbf{e^-}$ approaches the free volume an influence of optical modes starts to be negligible because of diminishing separation of charges of the approaching pair. The majority of phonon--exciton interaction is caused by lattice deformations producing longitudinal acoustic phonons. Assuming that an exciton can be described by Gaussian function of radial extent $a$
\begin{equation}
\psi(\mathbf{r})=(\sqrt{2}/a)^{3/2} \exp{[-\pi(r/a)^2]},
\end{equation}
after minimization procedure, the energy of a exciton-phonon system can be expressed as
\begin{equation}
E(\lambda)=B[\lambda^2-g_s \lambda^3-g_1 \lambda],
\end{equation}
where $B=3\pi\hbar^2/(2m a_0^2)$, where $m$-- an exciton mass, $a_0$-- a lattice constant, and $\lambda=a_0/a$, whereas $g_s$ and $g_1$ are acoustic and optical phonon coupling parameters respectively. The magnitude of $g_s$ is a crucial factor which sets at which $\lambda$ the energy of phonon-exciton interaction is negative (self-trapped exciton); the larger $g_s$ the less localised excitons are trapped into the acoustic phonons potential.\\
Because $g_s$ is a function of coordination number $\nu$
\begin{equation}
g_s=E_d^2/(2M\nu V u^2),
\end{equation}
where $E_d$-- deformation potential, $u$-- velocity of sound, $M$ and $V$-- mass and volume of a unit cell, respectively, its magnitude diminishes at an edge of a free volume. If yes, the phonon-exciton interaction is more attractive for even pretty delocalised exciton (an attractive interaction appears for lesser $\lambda$ and a depth of a minimum deepens). One can deduce that the transition into a free volume requires an intake of energy to break an interaction of the pair with acoustic phonons. In this case (let call the transition to a cavity as $\epsilon \rightarrow \epsilon_0$ due to a permittivity change during the transition) an excess of energy related to the transition $\epsilon \rightarrow \epsilon_0$ could compensate the interaction with phonons. The balance of these two pieces of energy varies for different materials and is characterised by their $g_s$.\\
Yet, it was assumed that $\mathbf{e^+}$--$\mathbf{e^-}$ enters a free volume as a pair. The Onsager radius for this pair is ca.~135$\AA$ for $\epsilon$=2. For larger distances the motion of each particle is an uncorrelated random walk but a diameter of a blob is not larger \cite{Stepanov03}. If one assumes that a typical alkane has 20 carbon atoms (the length of each C-C segment is 1.5~$\AA$), and that a typical free volume distance can be approximated by interlamellar distances 30~$\AA$, the initial distance of one of the charges from the free volume of some angstroms could be less than that from the partner (it is highly probable that a blob volume of, say, diameter 80~$\AA$ does overlap with a free volume(s)). Thus, the positron or electron from a blob can reach likely a free volume before it forms a tightly bond pair with a partner. In this case however, when a transition $\epsilon\rightarrow\epsilon_0$ takes place, entering the partner into the free volume is in a radial direction (perpendicularly to a wall border). Thus, a momentum change is radial and cannot be transferred by phononic modes which are longitudinal only (one assumes an inflexible border here; a lattice forms connections along a border of free volume and cannot oscillate in a perpendicular direction). To summarise, if one of the partners enters a free volume first, it is not possible to transfer by phonons a momentum excess caused by a transition $\epsilon\rightarrow\epsilon_0$. In this case a photonic transition can satisfy it. To generalise, one may conclude that a competition of phononic and photonic transitions is a function of $g_s$, $\epsilon_r$, a density and radii of free volumes and an ionisation energy of the sample material.
\paragraph{Oscillator strength for transition from the bulk to the free volume.}
Although the process of photonic deexcitations is common for excitons it seems to be neglected in the case of positron studies yet. However, in cases when selection rules for the optical transition are respected the emission of the photon should be taken into account.\\
As a simplified approach to this kind of transition, let us consider photonic transition of a loosely connected pair of particles $e^-$ and $e^+$ near the free volume (described by a wave function $\Psi_b$) to the state $\Psi_a$ describing the pair in the free volume. Let us assume a free volume large enough to contain a Ps atom for which the influence of the wall can be neglected. Furthermore, $\Psi_b$ is approximated by the state for which the influence of the free volume's boundary can be neglected. These simplifications could replace the case where one considers stationary states which are solutions of (\ref{eq:Hamiltonian}) for the finite quantum well. In the latter case, photonic deexcitations are considered in the exciton theory. Instead of this, let us consider the photonic transition between the $\Psi_i$ states in the simplified model specified above.\\
According to the hydrogenic model, the motion of the pair around the center of mass can be related to some angular momentum--an angular momentum operator commutes with the Hamiltonian (\ref{eq:Hamiltonian}). The transition to the free volume with photonic emission should be connected with a change in the angular momentum\footnote{ The angular momentum of $\Psi_b$ should be recalculated from the center-of-mass coordinate system to a coordinate system related to a center of free volume. However, the shift is negligible here. Using the approximation from classical mechanics, this difference is two orders less than the value of an angular momentum in a relative motion in the hydrogen-like atom, $\hbar$. Furthermore, the correction due to the thermal motion of the center of mass around a free volume is to be neglected, too.}. The selection rule for a dipole transition $L=1 \rightarrow L=0$ is $\Delta L$=-1, \cite{Fox01}.\\
The probability of photonic transition between states is expressed by oscillator strength \cite{Fox01}. The nearly universal model of calculations of oscillator strength in the case of an exciton in a quantum dot was given in \cite{Takagahara93}. With the support of this work, the oscillator strength for transition from the state $\Psi_b$ to the hydrogen-like state $\Psi_a$, with $n$=1, $L=0$ can be calculated in a dipole approximation (It is assumed that the free volume is large enough to contain the $n$=1 state of $Ps$).\\
The photonic transition rate from $\Psi_b$ to $\Psi_a$ is proportional to
\begin{equation}
\int_\mathbf{K} d\mathbf{K}\ |M_{ba}|^2,
\label{eq:OscillatorStrength1}
\end{equation}
where
\begin{equation}
M_{ba}=\int d\boldsymbol{\varepsilon}_0\ \langle\Psi_a|e\ \mathbf{r}\cdot\boldsymbol{\varepsilon}_0|\Psi_b\rangle,
\label{eq:MatrixElement}
\end{equation}
where $e$ -- an electron charge and $\mathbf{r}$ -- spatial coordinates. The latter integral is calculated over all polarization $\boldsymbol{\varepsilon}_0$ orientations (should not be confused with a vacuum permittivity $\epsilon_0$), whereas the band-like indices $\mathbf{K}$ (relating to the exciton theory nomenclature) in the integral (\ref{eq:OscillatorStrength1}) corresponds to this index in (\ref{eq:EnergyInBulk}).\\
Taking into account the simplification that the final state during the photonic transition is nearly the ground state of Ps in vacuum $\Psi_{1s}^{vac}$, and that the above-mentioned selection rule should be obeyed, the initial state should be an eigenstate $\Psi_{2p}^{bulk}$ mentioned in (\ref{eq:EnergyInBulk}).
\\
In our simplification, it can be assumed that $\Psi_a$ is an atom-like state with energy states given by (\ref{eq:EnergyInBulk}) and $V_i$=0. Furthermore, following a theory given in \cite{Takagahara93}, this state can be assumed to be
{\footnotesize
\begin{eqnarray}
\lefteqn{\Psi_{LM}^{(n)}(\mathbf{r}_e,\mathbf{r}_h)=C\ r_{eh}^n\ e^{-\alpha\ r_{eh}} \times {} }\\
& & \times \sum_{l_e,l_h,m_e} \langle l_e,m_e,l_h,m_h|L,M \rangle\ Y_{l_e,m_e}(\Omega_e) Y_{l_h,m_h}(\Omega_h) R^{(n)}_{l_e,l_h}(r_e,r_h),\nonumber
\end{eqnarray}
}
where $r_{eh}=|\mathbf{r}_e-\mathbf{r}_h|$, the $\langle l_e,m_e,l_h,m_h|L,M \rangle$ are standard Clebsh-Gordan coefficients and $Y_{lm}(\Omega_i)$ are spherical harmonics, whereas
{\footnotesize
\begin{equation}
R_{l_e,l_h}^{(n)}(r_e,r_h)=\sum_{l_e',l_h'}\sum_{z_e,z_h}C^{(n)}_{l_e,l_h}(l_e',z_e,l_h',z_h)\ j_{l_e'}\big( k_{z_e}^{l_e'}\frac{r_e}{R} \big)\ j_{l_h'}\big( k_{z_h}^{l_h'}\frac{r_h}{R} \big),
\end{equation}
}
is the radial part consisting of spherical Bessel functions $j_l(r_i)$ for which $l_e$, $l_h$ are special cases appearing in the $l_e'$, $l_h'$ set.\\
In our case, $n$=1, $L$=0 and we assume that the state $\Psi_a$ is a mixture of all states $M$=0,$\pm$1. Then the expression (\ref{eq:OscillatorStrength1}) can be replaced by
\begin{equation}
|\sum_{i=e,h;\ j=1\dots 3} \langle \Psi_a (\mathbf{r}_e,\mathbf{r}_h) | p_{i}^{j} |\Psi_b (\mathbf{r}_e,\mathbf{r}_h) \rangle|^2,
\label{eqn:Envelope}
\end{equation}
where $p_{i}^{j}$ are a momentum operator elements~\cite{Henry70,Bryant88}.\\
Of course, the above expressions are non-zero in general which means that the optical transition in these cases is possible.\\
The formula~(\ref{eqn:Envelope}) is motivated by the envelope function approximation~\cite{Harrison05,Coon84}. Here, atomic functions $\Psi$ represent the rapidly varying component whereas the envelope functions $\alpha_i$ could be approximated as Gaussian functions that overlap partially which is suggested by the result shown in fig.~\ref{fig:Kayanuma90}. More generally, the oscillator strength should be considered as proportional to
\begin{equation}
\langle\Psi_a|\boldsymbol{\varepsilon}_0\cdot\mathbf{p}|\Psi_b\rangle\ \langle \alpha_a|\alpha_b\rangle+
\langle\Psi_a|\Psi_b\rangle\ \langle \alpha_a|\boldsymbol{\varepsilon}_0\cdot\mathbf{p}|\alpha_b\rangle
\end{equation}
but the scalar product of the $\Psi$ functions in the last term of this equation is negligible (atomic states). The matrix element in the first part of this equation, when summed over all polarization vectors, is given by~eq.~(\ref{eqn:Envelope}).
\paragraph{Appendix: What is the relative probability that the $e^+$--$e^-$ pair in the bulk is in the $\Psi_{2p}$ state?}
Assuming $\epsilon_r$=2 (characteristic for molecular crystals with non-polar bonds, like alkanes) the kinetic energy of a thermalised particle at room temperature, 0.025~eV, equals to an absolute value of interaction energy for a hydrogen-like atom at a distance 135$\AA$ (Onsager radius\footnote{For a non-atomic character of the interaction of the pair (quasi-free particles), the energy at this distance is ca. 0.05~eV}). This means that for this distance and less, the particles start increasingly to approach one another. However, the energy of electrostatic attraction is more influential (one order greater) than the thermal one for the distance compared to $n$=4 in the atom ($a^{(4)}$=34$\AA$, $E^{(4)}$=-0.11eV). This means that the states with $n$=4 and less are produced with greater efficiency due to the more effective electrostatic attraction of the pair. Furthermore, when the electron is thermalised, its velocity at room temperature is ca.~$10^5$~m/s. This value is much lower than the relative velocity of an electron in the ground state of a hydrogen-like atom which is ca.~2$\cdot$10$^6$~m/s. Thus, only a small fraction of thermalised electrons and positrons are able to form an atom-like structure in the bulk without injection of energy from the surroundings. However, because the relative velocity of an electron in a hydrogen-like atom in excited states (e.g.:~$v^{(4)}\simeq 5\cdot10^5~m/s$) is comparable to the maximum of that from the relative velocity distribution related to thermal movement, the probability of creation of the excited state of the pair by thermalised particles is more probable.
\paragraph{Summary and conclusions}
It was shown that the calculation regime known from the exciton theory can be applied to develop a theory of positronium formation in matter. In particular, the oscillator strength for a transition from a quasi-free pair in the bulk to the positronium atom in a free volume has been approximated. The value of this expression is a measure of probability of photonic transition for the process of positronium formation.\\
Because the energy of the $e^+$--$e^-$ pair state in a bulk is related to the electronic properties of the bulk, detection of photonic radiation from deexcitation (its energy, polarization, etc.) could give new insight into matter, improving the positron spectroscopy.
\section{Acknowledgements}
The Author wants to thank Prof.~M.~Za{\l}u\.zny (M.~Curie-Sk{\l}odowska University, Lublin, Poland) for recommendation of the literature.\\
This work was supported by the grant 2013/09/D/ST2/03712 of the National Science Center in Poland.
\bibliography{PaperBiblio_Pietrow}{}
\bibliographystyle{unsrt}
\end{document}